\documentclass[article]{aa}

\usepackage{graphicx}
\usepackage{natbib}
\usepackage{amsmath}
\usepackage{mathrsfs}
\usepackage{xcolor}
\usepackage{txfonts}
\usepackage[colorlinks=true]{hyperref}
\hypersetup{linkcolor=blue,citecolor=blue,filecolor=blue,urlcolor=blue}
\usepackage[normalem]{ulem}

\begin{document} 

\title{Radiation-pressure instability is an artifact of constant-$\alpha$ closure}
\subtitle{Implications for AGN disk tensions}

\titlerunning{stable disk $\alpha$-prescription}

\author{M. H. Naddaf        \inst{1}
\and    M. Ghasemnezhad     \inst{2}
\and    H. Ghanbarnejad     \inst{2}
\and    D. Hutsem\'ekers    \inst{1}
\and    B. Czerny           \inst{3}}

\institute{
Institut d'Astrophysique et de Géophysique, Université de Liège, Allée du six août 19c, B-4000 Liège (Sart-Tilman), Belgium \and
Department of Interdisciplinary Physics and Technology, Faculty of Science, Shahid Bahonar University of Kerman, Kerman, Iran \and
Center for Theoretical Physics, Polish Academy of Sciences, Al. Lotnik\'ow 32/46, 02-668 Warsaw, Poland}
   
 
\abstract
{}
{The standard $\alpha$-disk formalism parametrizes turbulent angular momentum transport through a dimensionless coefficient $\alpha$, assumed to be spatially and thermodynamically invariant. While analytically convenient, this assumption leads to the well-known thermal and viscous instabilities in radiation-pressure dominated (RPD) regions. We show that this instability is not the consequence of radiation pressure, but is due to enforcing a constant $\alpha$ across distinct thermodynamic regimes.}
{Requiring the steady thin-disk (TD) to remain thermally stable and single-valued in the $\dot{M}$--$\Sigma$ plane yields a necessary condition on the stress response, expressed as $\eta_{\rm x} \equiv d\ln\alpha_{\rm x}\,/\,d\ln X > 4/7$, where $X \equiv P_{\rm gas}/P_{\rm rad}$.
The resulting constraint on $\alpha_{\rm x} \equiv \alpha(X)$ is obtained within the same height-integrated TD framework that produces the classical unstable branch: it is the stress response required for the steady branch to remain thermally stable and single-valued, while preserving the usual thin-disk regime conditions.}
{$\alpha_{\rm x}$ removes the RPD unstable branch. The disk structure becomes smooth and globally single-valued, with higher $\Sigma$ and $\tau$ in the inner RPD disk, while preserving the standard effective-temperature profile. This increases thermal and inflow timescales, offering a natural route to accretion-state dependent variability without large-amplitude radiation-pressure limit cycles. It also motivates revisiting AGN disk tensions, including microlensing sizes and continuum reverberation lags with improved radiative-transfer modeling. 
The results show that the classical RPD instability is tied to the constant-$\alpha$ closure, rather than to radiation pressure alone.}
{}

\keywords{accretion, accretion disks -- instabilities -- black hole physics -- radiation mechanisms: general -- methods: analytical}

\maketitle
\nolinenumbers

\section{Introduction}\label{intro}

The standard TD model of \citet{SS1973} remains the canonical framework for describing radiatively efficient accretion onto compact objects and underlies much of the standard interpretation of active galactic nuclei and black hole growth \citep{netzer2013}. Its central simplification is the replacement of the unknown transport physics by a local closure, in which the turbulent stress is set as a fraction of the local total mid-plane pressure through the dimensionless parameter $\alpha$. This prescription yields the familiar analytic structure equations and has underpinned much of modern accretion theory. Yet, it leaves unresolved the basic physical question of why the effective stress should remain independent of the local thermodynamic regime \citep{pringle1981}.
The thermal and viscous instability of radiation-pressure-dominated (RPD) regions when the stress scales with total pressure is the classical \citet{lightman1974} instability, later developed by \citet{ss1976} and generalized by \citet{piran1978}. In this case, the $\dot{M}$--$\Sigma$ equilibrium curve develops a negative-slope segment, producing the familiar unstable branch and motivating long-standing discussions of large-amplitude disk variability and limit-cycle behavior. The physical origin of this instability however remained uncertain, as constant $\alpha$ is not a prediction of disk microphysics.

In sufficiently ionized disks, angular momentum transport is expected to arise from MHD turbulence driven by magnetorotational instability (MRI), rather than from a universal hydrodynamic viscosity \citep{balbusHawley1991, hawley1995}. Thus, $\alpha$ should be regarded as an effective closure parameter describing the nonlinear saturation of MRI, which may depend on the disk thermodynamic state. There is thus no {\it a priori} reason for $\alpha$ to remain invariant across the transition from gas-pressure dominated (GPD) to RPD regimes.
Early magnetic-viscosity arguments already suggested that the inner RPD disk need not follow the naive total-pressure stress law \citep{Sakimoto1981}. GRMHD simulations likewise indicate that $\alpha$ need not be spatially constant \citet{Penna2013, Abramowicz2026}.

This is reinforced by empirical, observational, and theoretical studies of thin, fully ionized disks showing that $\alpha$ is not universal. Typical values are of order $\alpha\sim0.1$, while viable observational and numerical ones span $10^{-4} \lesssim \alpha \lesssim 0.3$ \citep{king2007,hirose2009}. More importantly, radiation-MHD simulations show that the RPD regime can deviate from constant-$\alpha$ models (hereafter SS models). Stratified shearing-box calculations by \citet{hirose2009} found no thermal runaway showing that stress fluctuations can precede total-pressure fluctuations, challenging the interpretation of stress as a local function of total pressure. Larger-box simulations recover runaway behavior, but with heating, cooling, and vertical energy transport largely different from the SS expectations \citep[e.g.,][]{jiang2013b, jiang2019}. The issue is thus whether a rigid constant-$\alpha$ closure adequately captures RPD MRI turbulence \citep{blaes2025}.

Proposed remedies include gas (or mixed) pressure scaling in the stress law \citep{Sakimoto1981, Stella1984, merloni2006, grzedzielski2017}; magnetically supported disks with altered vertical structure and pressure balance \citep{begelman2007, zheng2011, sadowski2016a}; incorporating wind-driven mass-loss that changes the local $\dot{M}$ and $\Sigma$ structure \citep{Poutanen2007, Laor2014, habibi2019}; or adopting slim-disk solutions in which advective cooling becomes important \citep{abramowicz1988, Chen1995}. These approaches achieve stability by modifying the stress law, invoke additional physics, or leave the standard TD branch.

Our approach is deliberately minimal. We retain the standard height-integrated, radiatively efficient TD equations and ask what effective stress response to $X\equiv P_{\rm gas}/P_{\rm rad}$ is required for the steady $\dot{M}$--$\Sigma$ branch to remain thermally stable and single-valued. This is neither a first-principles MRI-saturation model nor an independent phenomenological guess: the condition is derived at the same TD level as the classical \citet{lightman1974} unstable branch. Once the required response is satisfied, the unstable branch disappears while the solution remains geometrically thin and optically thick.

The paper is organized as follows. In Section~\ref{sec:prescription} we introduce the consistent $\alpha_{\rm x}$ closure derived from both the local thermal stability criterion and the structure of the $\dot{M}$--$\Sigma$ equilibrium curve. In Section~\ref{sec:results} we show how it removes the classical unstable branch and modifies the disk structure. It is followed by discussion in Section~\ref{sec:discussion}. We summarize the implications in Section~\ref{sec:conclusion}.

\section{Self-consistent $\alpha_{\rm x}$ prescription}
\label{sec:prescription}

We next examine how $\alpha_{\rm x}$ affects the TD structure. We use the classical height-integrated, steady-state \citet{SS1973} framework, adopting Newtonian gravity for simplicity. The only new element is that $\alpha$ is no longer constant, but depends locally on the thermodynamic state via the local pressure partition $X$, as
\begin{equation}
\alpha_{\rm x} \equiv \alpha(X),
\qquad \qquad {\rm where} \qquad \qquad X \equiv {P_{\rm gas}}/{P_{\rm rad}}.
\end{equation}

Any change in $X$ is thus reflected in value of $\alpha$, i.e.
\begin{equation}\label{eq:deltalnalpha}
\delta \ln \alpha = \eta_{\rm x}\,\delta \ln X,
\end{equation}
or, equivalently,
\begin{equation}
\eta_{\rm x} = {d\ln\alpha}\,/\,{d\ln X}.
\end{equation}

Here $\alpha_{\rm x}$ is a reduced, height-integrated effective closure, not a microscopic MRI-saturation law. If magnetic stresses contribute to angular momentum transport, they are not introduced as an additional vertical-support term in the present TD structure; their effect is absorbed into the effective stress coefficient. We use this closure to find the stress response required for the steady TD branch to remain thermally stable and single-valued, which gives $\eta_{\rm x}>4/7$, as derived in the following.

\subsection{Basic equations}

The following fundamental equations describe a steady, Keplerian, optically thick structure.

(a) Vertical hydrostatic equilibrium:
\begin{equation}
P_{\rm tot} = \Omega_K^2 \Sigma H\, /2,
\label{eq:hydrostatic}
\end{equation}
where $\Omega_K = (GM/R^3)^{1/2}$ is the Keplerian angular velocity, and $\Sigma = 2\rho H$ is the surface density, where $\rho$ is the midplane density, and $H$ is the pressure scale height.
\begin{equation}
P_{\rm tot} = P_{\rm gas} + P_{\rm rad},
\end{equation}
is the total midplane pressure consisting of gas pressure and radiation pressure, where
\begin{align}
\label{eq:Pgas}
&P_{\rm gas} = (k_{\rm B}/m_{\rm H})~\Sigma\, T_c\, H^{-1},\\
\label{eq:Prad}
&P_{\rm rad} = a~T_c^4\, / 3,
\end{align}
where $a$ is the radiation constant, $k_{\rm B}$ the Boltzmann constant, and $m_{\rm H}$ is the hydrogen mass.

(b) Angular momentum transport
\begin{equation}
3\pi\, \nu \Sigma = \dot{M} \mathcal{J}(R),
\label{eq:angmom}
\end{equation}
where $\dot{M}$ is the accretion rate, $\nu = \alpha c_s H$ is the kinematic viscosity, $c_s=\Omega_K H$ is the sound speed, and $\mathcal{J}=1 - \sqrt{R_{\rm in}/{R}\,}$ is the inner boundary correction.

(c) Energy balance
\begin{equation}
Q_{\rm vis} = Q_{\rm rad} + Q_{\rm adv},
\end{equation}
i.e. viscous dissipation balances cooling though radiation and advection. These quantities are defined in the following.
\begin{equation}
Q_{\rm vis} = ({3}/{2})\,\Omega_K H \alpha P_{\rm tot}.
\end{equation}

The advective term is formally
\begin{equation}
Q_{\rm adv} =
\left(\frac{\dot{M}}{2\pi R}\right) T_c \frac{ds}{dR},
\end{equation}
and therefore depends on the radial entropy gradient. For reference, a local height-integrated estimate may be written as
\begin{equation}
Q_{\rm adv} \simeq
{\xi_{\rm adv}\dot{M}\Omega_K^2H^2}/({2\pi R^2}),
\end{equation}
where $\xi_{\rm adv}$ is an order-unity coefficient encoding the radial derivative terms. In this work, however, the stability bound is evaluated in the radiatively cooled RPD limit, where $Q_{\rm adv}\ll Q_{\rm rad}$. The result therefore does not rely on the algebraic estimate of $Q_{\rm adv}$.
\begin{equation}
 Q_{\rm rad} = 2 \times {16 \sigma T_c^4}/{3 \tau},
\end{equation}
where $T_c$ is the midplane temperature, $\tau=\kappa \Sigma/2$ is the vertical optical depth
and $\kappa$ is the Rosseland mean opacity. The factor 2 in $Q_{\rm rad}$ denotes the total radiative cooling from both disk faces.
Additional vertical heat transport by turbulence or magnetic buoyancy is not included explicitly; its effect would have to be absorbed into an effective cooling prescription beyond the scope of this work.

\subsection{Thermal stability analysis with $\alpha_{\rm x}$}

For an infinitesimal temperature perturbation while \(\Sigma\) is taken to be constant, thermal stability requires
\begin{equation}
\left({\partial}
\left[Q_{\rm vis}-Q_{\rm rad}-Q_{\rm adv}\right]/{\partial T}\right)_{\Sigma} < 0,
\end{equation}
or, since $T>0$, simply
\begin{equation}
\delta Q_{\rm vis}-\delta Q_{\rm rad}-\delta Q_{\rm adv}<0.
\end{equation}
Defining
\begin{equation}
f_{\rm a}\equiv {Q_{\rm adv}}/{Q_{\rm vis}},
~~
f_{\rm r}\equiv {Q_{\rm rad}}/{Q_{\rm vis}},
~~
f_{\rm r}+f_{\rm a}=1,
\end{equation}

Then
\begin{equation}
Q_{\rm vis}\left[\delta\ln Q_{\rm vis}
-f_{\rm r}\,\delta\ln Q_{\rm rad}
-f_{\rm a}\,\delta\ln Q_{\rm adv}\right]<0
\end{equation}

Since $Q_{\rm vis}>0$, the sign is set only by the bracket.

The logarithmic responses of viscous heating, diffusive radiative cooling, and advective cooling are
\begin{equation}\label{eq:deltaenergies}
\begin{split}
&\delta \ln Q_{\rm vis} = \delta \ln \alpha+\delta \ln P_{\rm tot}+\delta \ln H,\\
&\delta \ln Q_{\rm rad} = 4\,\delta \ln T,\\
&\delta \ln Q_{\rm adv} = \delta \ln \dot M+2\,\delta \ln H.
\end{split}
\end{equation}

In order to proceed, we define
\begin{equation}
\beta_{\rm r}\equiv {P_{\rm rad}}/{P_{\rm tot}},
~~\beta_{\rm g}\equiv {P_{\rm gas}}/{P_{\rm tot}},
~~\beta_{\rm r}+\beta_{\rm g}=1.
\end{equation}

Therefore, at a given radius for a fixed $\Sigma$,
\begin{equation}\label{eq:deltapressures}
\begin{split}
&\delta \ln P_{\rm tot}=\delta \ln H,\\
&\delta \ln P_{\rm tot}=\beta_{\rm r}\,\delta \ln P_{\rm rad}
+\beta_{\rm g}\,\delta \ln P_{\rm gas},\\
&\delta \ln P_{\rm gas}=\delta \ln T-\delta \ln H,\\
&\delta \ln P_{\rm rad}=4\,\delta \ln T,
\end{split}
\end{equation}
so
\begin{equation}\label{eq:deltalnH}
\delta \ln H= A\,\delta \ln T\,/D,
\end{equation}
where
$A \equiv 4-3\beta_{\rm g}$, and
$D \equiv 1+\beta_{\rm g}$.

Using $X \equiv P_{\rm gas}/P_{\rm rad}$, we have
\begin{equation}
\delta \ln X = \delta \ln P_{\rm gas} - \delta \ln P_{\rm rad},
\end{equation}
which, combined with Eqs. \ref{eq:deltalnalpha}, \ref{eq:deltapressures}, and \ref{eq:deltalnH}, yields
\begin{equation}
\delta \ln \alpha= -7\, \eta_{\rm x}~\delta \ln T\,/D
\end{equation}

Using the angular-momentum equation 
\begin{equation}
\begin{split}
&\delta \ln \dot M= \delta \ln \alpha + \delta \ln P_{\rm tot}+ \delta \ln H,\\
&\delta \ln \dot M= (2A-7\eta_{\rm x})\,\delta \ln T\,/D
\end{split}
\end{equation}

Now substituting into Eq.~(\ref{eq:deltaenergies}):
\begin{equation}
\begin{split}
&\delta \ln Q_{\rm vis}= (2A-7\eta_{\rm x})\,\delta \ln T\,/D,\\
&\delta \ln Q_{\rm adv}= (4A-7\eta_{\rm x})\,\delta \ln T\,/D.
\end{split}
\end{equation}

Then
\begin{equation}
\left[\frac{2A-7\eta_{\rm x}}D-4f_{\rm r}-f_{\rm a}\frac{4A-7\eta_{\rm x}}D\right]\delta \ln T<0.
\end{equation}

For a positive temperature perturbation, \(\delta\ln T>0\), and as $D$ is always positive, stability condition reads as
\begin{equation}
\Delta=2A-4D+4f_a(D-A)-7(1-f_{\rm a})\eta_{\rm x}<0,
\end{equation}

Since the classical \citet{lightman1974} instability is obtained in the radiatively cooled RPD branch, the critical bound is evaluated in the limit $\beta_{\rm g}\to0$ and $f_{\rm a}\to0$, thus
\begin{align}
&\Delta = 4-7 \eta_{\rm x}  < 0,\\
&\eta_{\rm x} > {4}/{7} \approx 0.571.
\end{align}

The result $\eta_{\rm x}>4/7$ therefore does not depend on adopting an algebraic approximation for $Q_{\rm adv}$.

The expression $\delta \ln Q_{\rm rad}=4\,\delta \ln T$ in Eq.~\ref{eq:deltaenergies} assumes locally temperature-independent opacity, appropriate to the electron-scattering-dominated RPD regime. If opacity variations are important, an additional $-\delta\ln\kappa$ term appears. This does not affect the bound $\eta_{\rm x}>4/7$, which is derived in the limiting RPD electron-scattering regime where the classical instability arises.

\subsection{Constraining $\alpha_{\rm x}$ from the $\dot{M}$--$\Sigma$ relation}

We now specialize to the RPD, radiatively cooled regime,
\begin{equation}
P_{\rm rad} \gg P_{\rm gas}, \qquad Q_{\rm adv} \ll Q_{\rm rad},
\end{equation}
in which the classical thermal–viscous instability arises. We derive a general constraint on the functional form of the viscosity $\alpha_{\rm x}$ from the structure of the $\dot{M}$--$\Sigma$ equilibrium curve, without assuming a specific parametric form. At fixed radius, the unstable branch lies in this regime, allowing the following scalings.

From vertical hydrostatic equilibrium,
\begin{equation}
P_{\rm tot} \propto \Sigma H,
\end{equation}
and in the radiatively cooled RPD limit,
\begin{equation}
P_{\rm tot} \approx P_{\rm rad} \propto T^4,
\end{equation}
which gives
\begin{equation}
H \propto {T^4}\,{\Sigma}^{-1}.
\end{equation}

The gas pressure scales as
\begin{equation}
P_{\rm gas} \propto \rho T \propto {\Sigma\, T}\, {H}^{-1},
\end{equation}
and therefore
\begin{equation}
P_{\rm gas} \propto {\Sigma^2}{T^{-3}}.
\end{equation}

Combining these expressions, we obtain
\begin{equation}
X \equiv {P_{\rm gas}}/{P_{\rm rad}} \propto {\Sigma^2}\,{T^{-7}}.
\end{equation}

The viscous heating rate scales as
\begin{equation}
Q_{\rm vis} \propto \alpha P_{\rm tot} H
\propto \alpha {T^8}{\Sigma}^{-1},
\end{equation}
while radiative cooling gives
\begin{equation}
Q_{\rm rad} \propto {T^4}{\Sigma}^{-1}.
\end{equation}

Thermal equilibrium $Q_{\rm vis} = Q_{\rm rad}$ then implies
$\alpha\,T^4 = \mathrm{const}.$, hence, $T^4 \propto \alpha^{-1}$.
Substituting into the expression for $X$ yields
\begin{equation}
X \propto \Sigma^2\,\alpha^{7/4},
\end{equation}
which determines $X(\Sigma)$ along the equilibrium branch

Taking the logarithm of the constraint equation,
\begin{equation}
\ln X - \frac{7}{4} \ln \alpha_{\rm x} = 2 \ln \Sigma + \mathrm{const},
\end{equation}
and differentiating with respect to $\ln \Sigma$, we obtain
\begin{equation}
{d\ln X}/{d\ln \Sigma} ={8}/{(4 - 7\eta_{\rm x})}.
\end{equation}

From angular momentum transport,
\begin{equation}
\dot{M} \propto \alpha P_{\rm tot} H \propto \alpha {T^8}{\Sigma}^{-1}.
\end{equation}

Using the thermal equilibrium condition, we obtain
\begin{equation}
\dot{M} \propto {\alpha_{\rm x}^{-1}}{\Sigma}^{-1}.
\end{equation}

Thus, the full $\dot{M}(\Sigma)$ relation is determined by the implicit
dependence $X(\Sigma)$.

Differentiating $\ln \dot{M}$ gives
\begin{equation}
{d\ln \dot{M}}/{d\ln \Sigma}
= -1 - \eta_{\rm x}~ {d\ln X}/{d\ln \Sigma}.
\end{equation}

Substituting, we find
\begin{equation}
{d\ln \dot{M}}/{d\ln \Sigma}
= ({\eta_{\rm x} + 4})/({7\eta_{\rm x}-4}).
\end{equation}

This equation gives the general slope of the $\dot{M}$–$\Sigma$ equilibrium curve for an arbitrary $\alpha_{\rm x}$. The classical instability corresponds to the denominator changing sign, producing a negative-slope segment. Requiring ${d\ln \dot{M}}/{d\ln \Sigma} > 0$, yields a necessary condition for the absence of the unstable branch. For $\eta_{\rm x} > 0$, this implies
\begin{equation}
\eta_{\rm x} > {4}/{7} \approx 0.571.
\end{equation}

\begin{figure*}[t]
    \centering
    \includegraphics[width=0.8\textwidth]{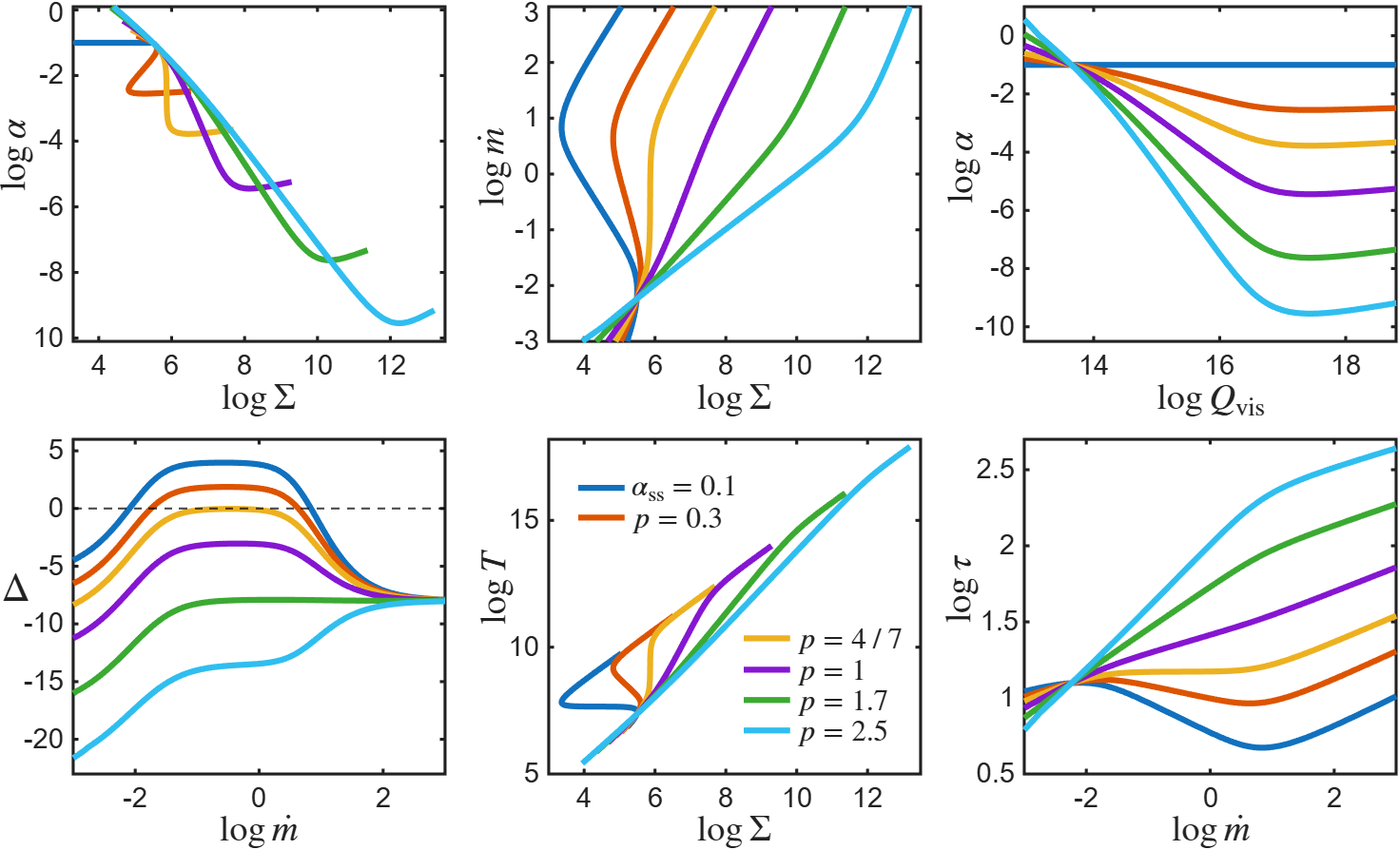}
    \caption{The disk thermal--viscous stability with regime-dependent $\alpha$, for $M_{\bullet}=10^8$ at $r=20$. Multi-panel view of steady-state TD solutions (in cgs units) computed with $\alpha_{\rm x}=\alpha_0\,X^{p}$, adopting $\alpha_0=0.1$ for illustration. Colors denote different indices $p$ as labeled; SS model ($p=0$) is labeled as $\alpha_{\rm ss}=0.1$. The equilibrium S-curves in the $\Sigma$--$\dot{M}$ plane (expressed in Eddington units, with $\dot{m}=\dot{M}/\dot{M}_{\rm Edd}$ and $\dot{M}_{\rm Edd}=L_{\rm Edd}/(\eta c^2)$ throughout the paper) is shown in \textit{top middle} panel; the characteristic middle (unstable) branch present for $\alpha_{\rm ss}$ flattens with increasing $p$ and disappears for $p \gtrsim 4/7$. $\Delta$ is the thermal stability diagnostic shown in \textit{bottom left} panel; the unstable regime ($\Delta>0$) progressively shrinks as $p$ increases and vanishes for $p \gtrsim 4/7$. \textit{Bottom middle} panel shows that the multi-valued structure of temperature solutions collapses to a single monotonic branch as the disk becomes globally stable. \textit{Bottom right} panel shows a monotonically increasing trend with $\dot{m}$ for stabilized disk, with significant enhancement of optical thickness (and thus $\Sigma$) in the inner disk, compared to standard SS model.}
    \label{fig:stability}
\end{figure*}

\subsection{Implications for $\alpha_{\rm x}$}

A most straightforward sufficient prescription satisfying the stability condition ($\eta_{\rm x} > {4}/{7}$), is a power-law form
\begin{equation}
\alpha_{\rm x}=\alpha_0\,X^{p}, \qquad p > 4/7
\end{equation}
where $\alpha_0$ is a normalization constant setting the overall amplitude of the effective viscosity.

Hence, the requirement of a single-valued, thermally stable equilibrium solution directly implies that the effective viscosity must increase sufficiently steeply with $X$. The key result is therefore the closure-level stability condition $\eta_{\rm x}>4/7$, with the power-law being only one sufficient compact realization of this more general response.

The power-law form should be interpreted as the local response of the effective viscosity to the pressure partition, rather than a global prescription for its absolute value. In practice, $\alpha$ remains bounded, since $X$ is finite across disk regimes, and the normalization $\alpha_0$ can be chosen to keep $\alpha$ within the empirically expected range. We adopt $X \equiv P_{\rm gas}/P_{\rm rad}$, so that RPD regions correspond to $X \ll 1$. This choice is arbitrary up to inversion: defining instead $P_{\rm rad}/P_{\rm gas}$ would lead to an equivalent constraint with reversed sign, without affecting the physical conclusion.

\section{Results}\label{sec:results}

Figure~\ref{fig:stability} shows how $\alpha_{\rm x}$ determines whether the classical unstable branch exists. It summarizes the behavior of disk solutions as a function of the slope $p$ in the emergent viscosity law $\alpha_{\rm x}\propto X^{p}$, for $M_{\bullet}=10^8$ at $r=20$, where $M_{\bullet}= M_{\rm bh}/M_{\odot}$ is the black hole mass in solar unit, and $r=R/R_g$ is the dimensionless radius with $R_g=GM_{\rm bh}/c^2$ the gravitational radius.

The top-left panel shows the effective $\alpha$ versus $\Sigma$. The larger the $p$ is, the more rapidly $\alpha$ declines, suppressing viscosity in RPD regimes. The global effect is shown in the top-middle panel. For small $p$, the classical S-shaped $\dot{M}$--$\Sigma$ relation of RPD TD is recovered \citep{SS1973, lightman1974}. As $p$ increases, the unstable middle branch shrinks and disappears for $p>4/7$; the sequence then becomes monotonic, removing the thermal-viscous instability. This agrees with earlier expectations that a strong decline in $\alpha$ in RPD regime suppresses the heating-cooling runaway \citep[e.g.,][]{Stella1984}, but here follows directly from TD equations. The top-right panel shows $\alpha$ versus viscous heating, revealing the weakened heating response that stabilizes the RPD branch. $\alpha$ falls steeply at low $Q_{\rm vis}$, then shows an inflection and flattens at high $Q_{\rm vis}$, reflecting changes in the dominant opacity and cooling mechanisms.
The bottom-middle panel shows the corresponding temperature profiles. For low $p$, the turning point of the unstable branch remains visible, whereas for larger $p$ the profiles become smooth and single-valued. Thus, the disappearance of the S-curve reflects a continuous thermodynamic structure rather than a jump between branches.
The bottom-left panel shows the thermal stability parameter $\Delta$. For small $p$, $\Delta$ becomes positive over a finite range of $\dot{M}$, recovering the classical instability criterion. As $p$ increases, the peak of $\Delta$ drops and eventually remains negative for all $\dot{M}$, demonstrating global thermal stability. The critical value $p=4/7$ thus emerges as the lower bound for marginal stability.
The bottom-right panel shows $\tau$ versus $\dot{m}$. For stabilized disks, $p > 4/7$, the optical depth increases monotonically with $\dot{m}$. At fixed $\dot{m}$ above $\sim -2~\rm dex$, larger $p$ gives systematically higher $\tau$, reflecting the stronger temperature and density dependence of the opacity. The curves also steepen at high $\dot{m}$ as the disk enters the RPD regime and the opacity approaches electron scattering.
Overall, Fig.~\ref{fig:stability} shows that shallow stress responses retain the classical unstable branch, whereas responses with $\eta_{\rm x}>4/7$ remove it and increase the optical depth, with $X$ acting as the effective closure variable governing RPD stability.

\section{Structural changes with $\alpha_{\rm x}$}
\label{sec:structure}

The total pressure can be written as
\begin{equation}
P_{\rm tot}=P_{\rm rad}+P_{\rm gas}=P_{\rm rad}(1+X)\propto T^4(1+X).
\end{equation}

From vertical hydrostatic balance,
\begin{equation}
 T^4\propto\, \Omega_K^{2} \, \Sigma H(1+X)^{-1},
\end{equation}
and from angular momentum transport,
\begin{equation}
\Sigma \propto \dot{M}
\mathcal{J}\, \alpha^{-1} \Omega_K^{-1} H^{-2}.
\end{equation}

Now using $X\propto {\Sigma}{H^{-1}T^{-3}}$, one obtains
\begin{align}
H^9 &\propto \dot{M} \mathcal{J}\, \Omega_K^{-7} \alpha^{-1} X^{-4} (1+X)^3,\\
T_c^9 &\propto \dot{M}^{2} \mathcal{J}^{2}\, \Omega_K^{4} \alpha^{-2} X (1+X)^{-3},\\
\Sigma^9 &\propto \dot{M}^{7} \mathcal{J}^{7}\, \Omega_K^{5}\, \alpha^{-7} X^{8} (1+X)^{-6}.
\end{align}

In the radiatively efficient TD limit, $Q_{\rm adv} \ll Q_{\rm rad}$, equating viscous heating with diffusive radiative cooling gives the leading-order thickness scaling
\begin{equation}
H \propto \kappa \dot{M} \mathcal{J} (1 + X).
\end{equation}
In the RPD limit, $X\ll1$, this scaling has no explicit dependence on $\alpha_0$ or $p$. The smallness of the resulting $H/R$ then provides an \textit{a posteriori} check that the stabilized branch remains geometrically thin.

Thus,
\begin{equation}
\label{eq:scalingX}
 X^{4} (1+X)^{6} \propto  \kappa^{-9}\,\dot{M}^{-8} \mathcal{J}^{-8}\, \alpha^{-1} \Omega_K^{-7}.
\end{equation}

At given $R$, $M$, and $\dot M$, we define
\begin{equation}
\delta\log Y\equiv \log Y_{\rm x}-\log Y_{\rm ss}
\end{equation}
as the logarithmic difference between the $\alpha_{\rm x}$ and constant-$\alpha$ solutions. This yields the following relations.
\begin{align}
\label{eq:Del_H1}
&\delta\ln H =\delta\ln \kappa + \delta \ln (1+X),\\
9\, &\delta \ln H=
-4\, \delta\ln X
+3\, \delta \ln (1+X)
- \delta\ln\alpha,\\
9\, &\delta\ln T=
\delta\ln X
-3\, \delta \ln (1+X)
-2\, \delta\ln\alpha,\\
9\, &\delta\ln\Sigma=
8\, \delta\ln X
-6\, \delta \ln (1+X)
-7\,\delta\ln\alpha,\\
\label{eq:Del_X}
4\,&\delta \ln X
+6\,\delta \ln (1+X)
=-9\, \delta\ln \kappa -\delta \ln \alpha.
\end{align}

Now in the following we compare the TD structures obtained with $\alpha_{\rm x}$ to those of the SS model using the differential logarithmic relations we just found (Eqs.~\ref{eq:Del_H1}--\ref{eq:Del_X}), focusing on the inner regions dominated by electron-scattering opacity, $\kappa=\kappa_{\rm es}$.

We define the logarithmic deviation from the SS model as
\begin{equation}
\label{eq:logdev}
\mathcal{A}=
-\delta \log \alpha =
\log \alpha_{\rm ss}
-\log \alpha_{\rm 0}
-p \log X_{\rm x},
\end{equation}
so we have
\begin{align}
6\, &\delta \log H=
~\, \mathcal{A}
-4\, \delta\log X,\\
6\, &\delta\log T=
~\, \mathcal{A}
+2\, \delta\log X,\\
3\, &\delta\log\Sigma=
2\, \mathcal{A}
+4\, \delta\log X,\\
4\, &\delta \log X
+6\,\delta \log (1+X)
=\mathcal{A},
\end{align}
and since
\begin{equation}
\log X_{\rm x}=\log X_{\rm ss}+\delta\log X,
\end{equation}
thus
\begin{equation}
(4+p)\delta \log X
+6\delta \log (1+X)
=\log (\alpha_{\rm ss}/\alpha_{\rm 0})
-p \log X_{\rm ss}.
\end{equation}

This relation connects the modified solution to the corresponding SS solution and is used below to obtain the limiting RPD and GPD deviations.

To evaluate the right-hand side, we use Eq.~\ref{eq:scalingX}, which gives
\begin{equation}
\label{eq:pressure_partition}
X^{4} \, (1+X)^{6} = C_0\, S^{-1}\, \alpha^{-1},
\end{equation}
where
$C_0 \simeq 10^{-23}
\left({\eta}/{0.1}\right)^8
\left({\kappa}/{0.34}\right)^{-9}
\left({\mu_m}/{0.615}\right)^{-4}$,
collects physical constants, and dimensionless parameter $S$ is given by
\begin{equation}
S={\dot{m}^{8}\,M_{\bullet}\,
\mathcal{J}^{8}(r)}\,{r^{-21/2}},
\end{equation}

This gives,
\begin{equation}
\log X_{\rm ss}= \left(\log C - \log S - \log \alpha_{\rm ss}\right)\,/q,
\end{equation}
where $q=4$ and $q=10$, in RPD ($X \ll 1$) and GPD ($X \gg 1$) regions, respectively.

From Eq.~\ref{eq:Del_H1}, we further note that in the RPD regime $\delta\log H \simeq 0$, but in the GPD regime $\delta\log H \simeq \delta\log X$. This shows that the $\alpha_{\rm x}$ prescription largely preserves the geometrically thin structure of the disk.

The structural scalings then reduce to
\begin{equation}
\label{eq:Arad}
\delta \log \Sigma =
4\, \delta\log T=
4\, \delta\log X=
\mathcal{A}_{\rm rad},
\end{equation}
in the RPD regime, and
\begin{equation}
\delta \log \Sigma =
4\, \delta\log T =
8\, \delta\log H =
8\, \delta\log X=
0.8\, \mathcal{A}_{\rm gas},
\end{equation}
in the GPD regime.

The parameter $\mathcal{A}$ can be written as
\begin{equation}
\mathcal{A}= \frac{p \log S
+(p+q) \log \alpha_{\rm ss}
-q \log \alpha_0
+23\, p}{(p+q)}.
\end{equation}

For $\alpha_{\rm ss}=0.1$, this becomes
\begin{equation}
\label{eq:deviation}
\mathcal{A}=
\frac{p}{(p+q)}\left( \log S + 22\right)
-\frac{q}{(p+q)} \left(\log \alpha_0+1\right).
\end{equation}

The resulting $\mathcal{A}$ provides an accurate description where the optically thick, radiatively cooled TD assumptions hold, that is $\log \dot{m} \gtrsim -1$.

\begin{figure}[t]
    \centering
    \includegraphics[width=\columnwidth]{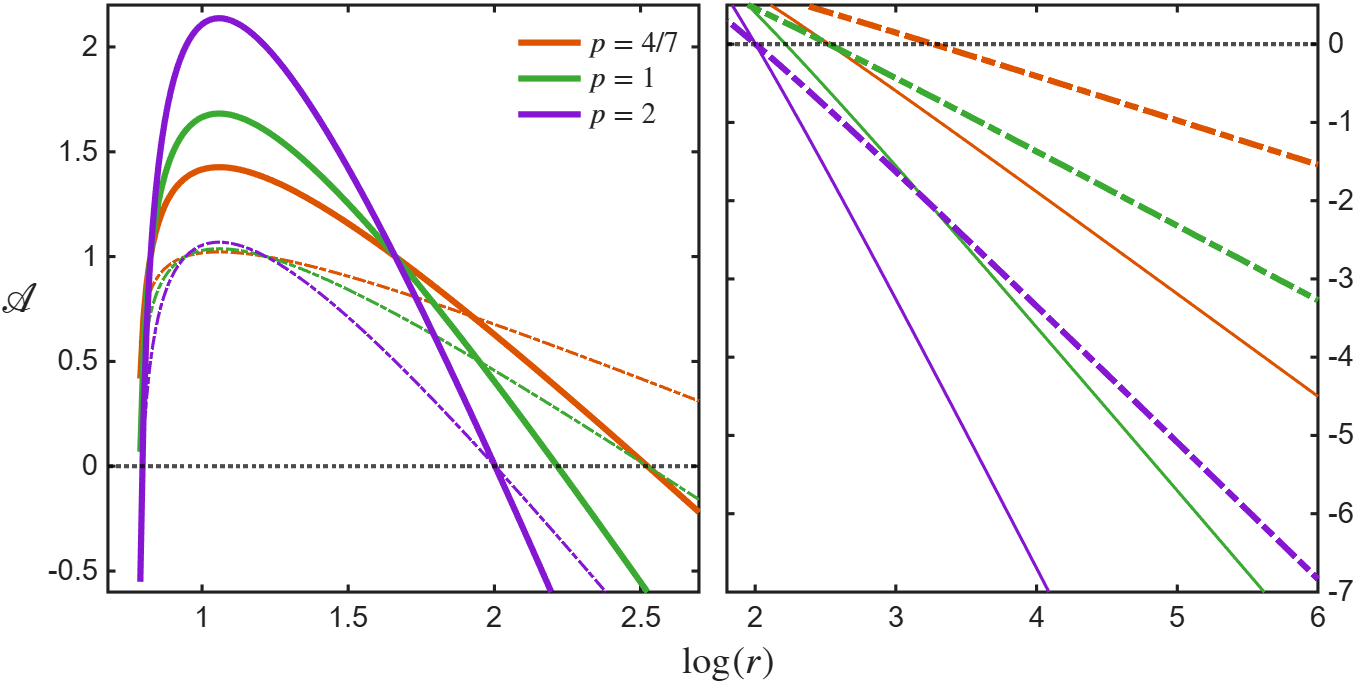}
    \caption{Radial behavior of the logarithmic deviation parameter $\mathcal{A}$ for $M_\bullet = 10^6$ and $\dot{m}=0.1$, with $\kappa=\kappa_{\rm es}$. Solid curves correspond to the RPD scaling ($\mathcal{A}_{\rm rad}$ with $q=4$), while dashed curves represent the GPD regime ($\mathcal{A}_{\rm gas}$ with $q=10$). The right panel extends the left panel. Different colors indicate the slope $p$ of the viscosity law, with $\alpha_0=0.01$. The horizontal dotted line marks $\mathcal{A}=0$, corresponding to no deviation from the standard SS solution.}
    \label{fig:comparison}
\end{figure}

Figure~\ref{fig:comparison} illustrates the radial behavior of $\mathcal{A}$ obtained from Eq.~\ref{eq:deviation} for representative $p$ values. The figure is intended as a visualization of the analytic scaling rather than as an independent numerical disk solution. In the RPD regime, $\mathcal{A}>0$ over the inner disk, corresponding to enhanced $\Sigma$ and therefore enhanced $\tau$ relative to the constant-$\alpha$ solution. Since $X$ increases outward (Eq.~\ref{eq:pressure_partition}), the disk gradually approaches the GPD regime, where $\mathcal{A}$ decreases and the solution tends back toward the standard branch.

The magnitude and radial extent of the deviation increase with $p$, reflecting the stronger sensitivity of the effective viscosity to the pressure partition. For the marginally stable case, $p=4/7$, the deviations are mild and extend over a broader radial range. Larger $p$ produces stronger but more localized deviations, yielding a sharper RPD--GPD transition as $\mathcal{A}$ declines rapidly outward. Where the dashed GPD curves cross below $\mathcal{A}=0$, the corresponding logarithmic changes become negative, indicating structural quantities smaller than in the constant-$\alpha$ solution.

Overall, the $\alpha_{\rm x}$ prescription preserves the leading-order thin-disk thickness while introducing controlled, regime-dependent changes mainly in $\Sigma$, $\tau$, and $T_c$. These changes remain continuous across the RPD--GPD transition, reflecting the removal of the multi-valued unstable branch.

\subsection{Asymptotic structural scaling solutions with $\alpha_{\rm x}$}

We now examine how the $\alpha_{\rm x}$ prescription modifies the asymptotic TD scalings.
The effective viscosity varies with the local pressure partition. This changes the response of disk structure to the system global parameters (Eq.~\ref{eq:scalingX}), as
\begin{equation}
X \propto
\left(\alpha_0\, \kappa^{9}\, \dot{M}^{8}\mathcal{J}^{8}\,\Omega_K^{7}
\right)^{-1/(q+p)},
\end{equation}
where $q=4$ and $q=10$, in RPD ($X \ll 1$) and GPD ($X \gg 1$) regions, respectively.

Thus, compared with the $\alpha_{\rm ss}$ case, the dependence of $X$ power on the system global parameters is weakened from $q^{-1}$ to $(q+p)^{-1}$.

The disk thickness in RPD regime, follows from substituting the asymptotic solution for $X$, yielding
\begin{equation}
\label{eq:Hrad}
H_{\rm rad} \propto \kappa\,\dot{M}\,\mathcal{J}.
\end{equation}

Therefore, in this limit, the explicit dependence on $\alpha_0$ and $p$ cancels from the leading-order thickness scaling. The main structural changes instead appear in the mid-plane temperature, and surface density, as
\begin{align}
T_{\rm rad} &\propto
\left(\kappa^{2p-1}\,\dot{M}^{2p}\mathcal{J}^{2p}\,\Omega_K^{2p+1}
\alpha_0^{-1}\right)^{1/(4+p)},\\
\Sigma_{\rm rad} &\propto
\left(\kappa^{7p-8}\,\dot{M}^{7p-4}\mathcal{J}^{7p-4}\,\Omega_K^{6p-4}
\alpha_0^{-4}\right)^{1/(4+p)}.
\end{align}

Thus, in the RPD branch, the $\alpha_{\rm x}$ prescription mainly redistributes the classical dependence of $T_c$ and $\Sigma$ on the disk system global parameters and on the viscosity normalization $\alpha_0$. Since $X\ll 1$, a positive $p$ implies $\alpha_{\rm x}=\alpha_0X^p < \alpha_0$, leading to a reduced effective viscosity in the RPD part of the disk.

In the opposite, GPD limit, we end up with
\begin{align} 
H_{\rm gas} &\propto \left(\kappa^{p+1}\,\dot{M}^{p+2}\,\mathcal{J}^{p+2}\,\Omega_K^{-7}
\alpha_0^{-1}\right)^{1/(10+p)},\\
T_{\rm gas} &\propto
\left(\kappa^{2p+2}\dot{M}^{2p+4}\mathcal{J}^{2p+4}\,\Omega_K^{2p+6}
\alpha_0^{-2}\right)^{1/(10+p)},\\
\Sigma_{\rm gas} &\propto
\left(\kappa^{7p-2}\,\dot{M}^{7p+6}\mathcal{J}^{7p+6}\,\Omega_K^{6p+4}
\alpha_0^{-8}\right)^{1/(10+p)}.
\end{align}

These asymptotic relations show explicitly where the new closure departs from the classical $\alpha_{\rm ss}$ solution. The modification does not simply rescale the SS disk by a constant factor; rather, it changes the local power-law response of the disk structure through the pressure partition ratio.

We emphasize that the usual TD regime checks are not violated by this closure. The scaling $H/R \propto \dot{m}\, \mathcal{J}\, r^{-1}$ is independent of $\alpha_0$ and $p$. The principal structural change is instead the increase of $\Sigma$, and therefore of $\tau$. The $\alpha_{\rm x}$ closure therefore removes the unstable low-$\Sigma$ branch while keeping the solution geometrically thin and optically thick.

The dimensionless form of asymptotic scaling relations derived under the $\alpha_{\rm x}$ prescription, separated into three main regimes analogous to the standard SS-disk regions, are presented in the following

\subsubsection{Region A: RPD ($X\ll1$), with $\kappa_{\rm es}$}
\vspace{-6pt}

\begin{alignat}{9}
H~ &\propto ~~\dot{m}\, \mathcal{J}\, M_{\bullet}\\
{\scriptstyle{H/R}} &\propto ~~\dot{m}\, \mathcal{J}\, r^{-1}\\
T_c ~&\propto \big(~M_{\bullet}^{-1} &&(~\dot{m}\mathcal{J}~)^{2p}~~~~~ &&r^{-3p-3/2}~~ &&\alpha_0^{-1}~\big)^w\\
\Sigma ~&\propto \big(~M_{\bullet}^{p} &&(~\dot{m}\mathcal{J}~)^{7p-4} &&r^{-9p+6} &&\alpha_0^{-4}~\big)^w\\
\rho ~&\propto \big(~M_{\bullet}^{-4} &&(~\dot{m}\mathcal{J}~)^{6p-8} &&r^{-9p+6} &&\alpha_0^{-4}~\big)^w\\
P ~&\propto \big(~M_{\bullet}^{-1} &&(~\dot{m}\mathcal{J}~)^{2p} &&r^{-3p-3/2} &&\alpha_0^{-1}~\big)^{4w}\\
X ~&\propto \big(~M_{\bullet}^{-1} &&(~\dot{m}\mathcal{J}~)^{-8} &&r^{21/2} 
&&\alpha_0^{-1}~\big)^{w}
\end{alignat}
with $w=1/(4+p)$.

\subsubsection{Region B: GPD ($X\gg1$), with $\kappa_{\rm es}$}
\vspace{-9pt}
\begin{alignat}{9}
H~ &\propto \big(~M_{\bullet}^{p+9}~~~ &&(~\dot{m}\mathcal{J}~)^{p+2}~~~~&&r^{21/2}~~~~~~~~ &&\alpha_0^{-1}~\big)^w\\
{\scriptstyle{H/R}} &\propto \big(~M_{\bullet}^{-1} &&(~\dot{m}\mathcal{J}~)^{p+2}&&r^{1/2-p} &&\alpha_0^{-1}~\big)^w\\
T_c ~&\propto \big(~M_{\bullet}^{-2} &&(~\dot{m}\mathcal{J}~)^{2p+4}&&r^{-3p-9} &&\alpha_0^{-2}~\big)^w\\
\Sigma ~&\propto \big(~M_{\bullet}^{p+2} &&(~\dot{m}\mathcal{J}~)^{7p+6}&&r^{-9p-6} &&\alpha_0^{-8}~\big)^w\\
\rho ~&\propto \big(~M_{\bullet}^{-7} &&(~\dot{m}\mathcal{J}~)^{6p+4}&&r^{33/2-9p} &&\alpha_0^{-7}~\big)^w\\
P ~&\propto \big(~M_{\bullet}^{-9} &&(~\dot{m}\mathcal{J}~)^{8p+8}&&r^{15/2-12p} &&\alpha_0^{-9}~\big)^w\\
X ~&\propto \big(~M_{\bullet}^{-1} &&(~\dot{m}\mathcal{J}~)^{-8} &&r^{21/2} 
&&\alpha_0^{-1}~\big)^{w}
\end{alignat}
with $w=1/(10+p)$.

\subsubsection{Region C: GPD ($X\gg1$), with $\kappa_{\rm ff}$}
In this region, the opacity is dominated by bound-free/free-free absorption (Kramers opacity) which has the form $\kappa_{\rm ff}\propto \rho\,T^{-7/2}$.
\begin{alignat}{12}
H~ &\propto &&\big(~M_{\bullet}^{p+9}~ &&(\dot{m}\mathcal{J})^{(p+3)/2}~~ &&r^{3(p+15)/4}~~~~ &&\alpha_0^{-1}~\big)^w\\
{\scriptstyle{H/R}} &\propto &&\big(~M_{\bullet}^{-1} &&(\dot{m}\mathcal{J})^{(p+3)/2}&&r^{-(p-5)/4} &&\alpha_0^{-1}~\big)^w\\
T_c ~&\propto &&\big(~M_{\bullet}^{-2} &&(\dot{m}\mathcal{J})^{p+3}&&r^{-3(p+5)/2} &&\alpha_0^{-2}~\big)^w\\
\Sigma ~&\propto &&\big(~M_{\bullet}^{p+2} &&(\dot{m}\mathcal{J})^{7(p+2)/2}&&r^{-15(p+2)/4} &&\alpha_0^{-8}~\big)^w\\
\rho ~&\propto &&\big(~M_{\bullet}^{-7} &&(\dot{m}\mathcal{J})^{3p+11/2}&&r^{(-18p-75)/4} &&\alpha_0^{-7}~\big)^w\\
P ~&\propto &&\big(~M_{\bullet}^{-9} &&(\dot{m}\mathcal{J})^{4p+17/2}&&r^{-6p-105/4} &&\alpha_0^{-9}~\big)^w\\
X ~&\propto &&\big(~M_{\bullet}^{-1} &&(~\dot{m}\mathcal{J}~)^{-7/2} &&r^{15/4} 
&&\alpha_0^{-1}~\big)^{w}
\end{alignat}
with $w=1/(10+p)$.

The classic SS asymptotic behavior is recovered for $p=0$, and the TD become globally stable for $p>4/7$.

It also reduces to the prescription of \citet{Sakimoto1981} for $p=1$, because in the RPD regime, $P_{\rm rad}\gg P_{\rm gas}$, so $P_{\rm tot}\simeq P_{\rm rad}$, and therefore $T_{r\phi}\propto \alpha_0 P_{\rm gas}.$
The key point, however, is that unlike \citet{Sakimoto1981}, who directly imposed gas-pressure scaling, the present closure, $\alpha_{\rm x}=\alpha_0 X^p$, follows from the closure-level stability condition and recovers a gas-pressure-like stress only as the $p=1$ RPD-limit case.

\section{Discussion}\label{sec:discussion}

\subsection{Implications for accretion studies}

The original radiation-pressure instability was derived within the SS framework \citep{lightman1974, ss1976}, with its presence depending on the adopted heating and cooling closures \citep{piran1978}. Our criterion is derived at the same level: within the radiatively cooled, height-integrated TD equations, we identify the stress response to $X$ required for the $\dot{M}$--$\Sigma$ branch to remain thermally stable and single-valued. If turbulent, convective, magnetic, or non-local heat transport is important, the detailed criterion may change; this caveat applies equally to the classical constant-$\alpha$ instability.

This differs from earlier stabilization strategies, which modify the stress law, invoke additional physics, or leave the standard TD branch. By contrast, we show that, even within the standard SS equations, the constant-$\alpha$ prescription is the specific reduced closure that generates the unstable RPD branch.

The MRI context sets both the motivation and the main calibration test of this approach. Since angular momentum transport in ionized disks is MRI-driven \citep{balbusHawley1991,hawley1995}, $\alpha$ should be viewed as a saturated, vertically averaged effective stress, not a fundamental constant. Numerical studies often report MRI transport in terms of magnetic pressure, net magnetic flux, or plasma beta, with stress increasing as magnetization increases \citep[e.g.,][]{hawley1995,salvesen2016}. These scalings are not disregarded; rather, in the present gas-plus-radiation TD branch they are absorbed into the coefficient $\alpha_{\rm eff}=W_{R\phi}/(P_{\rm gas}+P_{\rm rad})$. The stability result states that this total-pressure-normalized stress response must satisfy
\begin{equation}
\eta_{\rm x}^{\rm MRI}~\equiv~
~{d\ln\alpha_{\rm eff}}~/~{d\ln X}
~>~{4}/{7}
\end{equation}
to remove the classical radiatively cooled RPD unstable branch. If magnetic pressure instead becomes part of the vertical support, $P_{\rm mag}\sim P_{\rm gas}+P_{\rm rad}$, the disk has moved to a magnetically supported branch, which is a complementary stabilization route \citep[e.g.,][]{sadowski2016a,Mishra2020,scepi2024}. Thus, the present result should be read as a closure-level stability condition for gas-plus-radiation TD disks, and as a direct calibration target for radiation-MHD simulations.

Future work should connect this framework to advective, quasi-spherical, outflowing, and magnetically influenced accretion regimes \citep{Narayan_1994, abramowicz1995, king2007, Mosallanezhad2014}. AGN extensions should include opacity effects, especially the iron-opacity bump, time-dependent evolution with $\alpha_{\rm x}$ across RPD and unstable zones, and direct comparisons with 3D radiation-MHD simulations \citep[e.g.,][]{jiang2016, jiang2019}.

\subsection{Implications for AGNs}

Observed AGN disks show tensions with the simplest TD predictions, but the main implication of this closure is more basic: an RPD disk need not become low-column, thermally unstable, advective, or geometrically inflated. If the stress responds to the local thermodynamic state, the disk can remain RPD, optically thick, geometrically thin, and thermally stable. The closure does not remove radiation pressure; it changes the disk response to it.

The structural reason is direct. In RPD regions, where $X$ is small, $\alpha_{\rm x}$ is reduced. At fixed $\dot{m}$, it is thus compensated by raising $\Sigma$, rather than moving onto the low-column RPD branch. This removes the unstable low-$\Sigma$ solution and enhances the inner disk optical thickness as (see Eqs.~\ref{eq:logdev},
\ref{eq:Arad}, and \ref{eq:deviation})
\begin{equation}
\big( \delta \log \Sigma =
4\, \delta\log T=
4\, \delta\log X\big) =
~-\delta\log \alpha ~
\propto~ \log S \pm \mathcal{C}
\end{equation}
in the RPD regime where $S \propto \dot{m}^{8}M_\bullet{\cal J}^{8}(r)r^{-21/2}$, thus the enhancement grows with both $M_\bullet$ and, more strongly, $\dot{m}$.

Since $\alpha_{\rm x}$ reduces the sensitivity of heating to the RPD state and increases $\Sigma$ and $\tau$, it thus produces a more slowly evolving inner flow. In the leading RPD limit, the thickness scaling $H/R$ is nearly unchanged (see Eq.~\ref{eq:Hrad}), so the viscous, or equivalently inflow, timescale increases approximately as $t_{\rm inflow}\propto\alpha_{\rm x}^{-1}$. The thermal time follows the same dependence by definition.

\begin{figure*}[t]
    \centering
    \includegraphics[width=\textwidth]{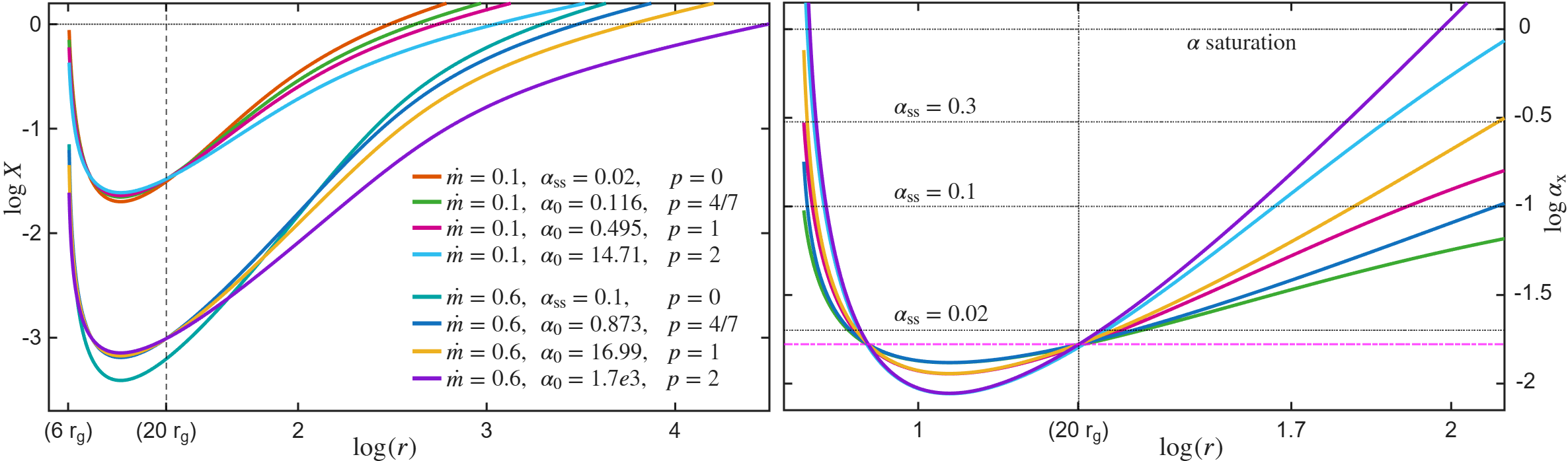}
    \caption{Radial behavior of $X$, and $\alpha^{-1}_{\rm x}$ for $M_{\bullet}= 10^9$. The magenta dash-dotted line marks a characteristic variability timescale of $300$ days at $20~r_{\rm g}$. The black horizontal dotted lines as annotated indicate different values of $\alpha_{\rm ss}$, and the lower horizontal line indicates the viscosity saturation threshold. Also, the opacity is treated as $\kappa=\kappa_{\rm es}+\kappa_{\rm ff}$, where $\kappa_{\rm ff}=\kappa_0\, \rho\,T^{-7/2}$, with $\kappa_0= 5\times10^{24}~~{\rm cm^5\, g^{-2}\, K^{7/2}}$ adopted here.}
    \label{fig:variability}
\end{figure*}

Therefore, $\delta\log t_{\rm thermal} \simeq
\delta\log t_{\rm inflow} \simeq
- \delta\log \alpha$,
which is positive over a broad range of inner disk radii, as shown in Fig.~\ref{fig:comparison}, and is also reflected in Fig.~\ref{fig:variability}.

This weakens the need for an early transition to a slim-disk at moderate to high $\dot{m}$. Stability can be obtained while remaining close to a TD-like branch, which is relevant for quasars, NLS1s, SEAMBHs, and high-redshift luminous sources whose optical/UV continua are often interpreted with TD-like scalings even at high inferred accretion rates \citep{Done2007, wang_shielding2014, DuPu2018, czerny2019Univ}. A suggestive example is the variable spectrum of a $z=6.51$ quasar accreting at $\lambda_{\rm Edd}\sim0.6$, consistent with $\lambda F_\lambda\propto\lambda^{-4/3}$  \citep{Leung2026}. Unlike slim disks, which stabilize high-accretion flows through advection in a geometrically thicker configuration \citep{abramowicz1988}, the present closure stabilizes the RPD branch through the stress response while preserving a thin, smooth $H/R$ structure.

Concerning variability, the classical RPD instability predicts large-amplitude burst--quench cycles, yet such cycles are rare in luminous AGNs, although they remain relevant for some extreme Galactic accretors such as GRS~1915+105 \citep{Belloni1997, Janiuk2011, Sniegowska2023}. The expected phenomenology of $\alpha_{\rm x}$ is, however, an optically thick, slowly evolving inner disk, rather than a violently intermittent one. This is qualitatively consistent with optical quasar variability studies, where damping times, variance, and power spectra depend on $M_\bullet$, wavelength, $\lambda_{\rm Edd}$, and the probed timescale \citep{McHardy2006, Arevalo2006, Kelly2009, Stone2022}. The observed suppression of optical variance with increasing $M_\bullet$ and $\lambda_{\rm Edd}$ provides an empirical benchmark for the predicted increase of thermal and inflow timescales in RPD disks \citep{Arevalo2023}. If the optical damping time traces the local thermal response, then $t_{\rm th}/t_{\rm dyn}\sim\alpha_{\rm x}^{-1}$. For $M_\bullet=10^9M_\odot$ at $r=20$, $t_{\rm dyn}\sim5$ days, so a characteristic timescale $\sim300$ days implies $\alpha_{\rm x}\sim0.017$, as shown in Fig.~\ref{fig:variability}.

Because $\alpha_{\rm x}\propto X^p$ and $X$ varies with radius, the closure can naturally produce a radial effective-viscosity profile too, providing an analytic analog of empirical or simulation-calibrated $\alpha(r)$ models \citep[see e.g.,][]{Penna2013, Abramowicz2026}.

The geometric implications are also important. The fraction of coronal radiation intercepted by the disk, BLR illumination, self-shadowing, and wind shielding all depend on $H/R$. Slim-disk self-shadowing models predict strong anisotropy and multiple BLR zones with different lags and line widths \citep{wang_shielding2014, DuPu2018}, while line-driven wind and obscurer models depend sensitively on geometry through X-ray shielding and UV illumination \citep{Murray1995, Proga2000, Giustini2023}. In the present model, radiation pressure does not by itself force the inner disk into a puffed-up funnel. A smoother, thinner $H/R$ profile thus implies less extreme irradiation and shielding geometry than in strongly inflated slim disks \citep{wang_shielding2014, DuPu2018}, while remaining closer to the thin branch than magnetically elevated or toroidally supported disks \citep{begelman2007, Begelman2017}.

The same structural changes may affect continuum-emitting regions. Microlensing and continuum-lag studies often infer optical sizes larger than predicted by the simplest TD model, while broadly preserving disk-like wavelength scalings \citep{Morgan2010, Jha2022, Hutsemekers2025}. Because the present height-integrated model preserves the effective-temperature profile, it does not by itself solve the blackbody size discrepancy. Its robust implication is instead structural: larger $\Sigma$ and $\tau$ can modify thermalization depths, color corrections, surface-brightness weighting, and lag normalization. This motivates revisiting such tensions with closure-updated disk-atmosphere and radiative-transfer calculations.

\section{Conclusions}\label{sec:conclusion}

We have shown that, within the standard radiatively cooled TD framework, removing the classical RPD unstable branch requires the effective stress coefficient to respond to the gas-to-radiation pressure partition. The unstable branch is thus not intrinsic to radiation pressure itself, but to imposing a thermodynamically invariant constant-$\alpha$ closure across distinct regimes. No additional dynamical branch, wind prescription, magnetic support term, or advective cooling solution is introduced. Instead, within the TD equations, the effective stress coefficient is allowed to respond to the local gas-to-radiation pressure partition, softening the RPD heating response and restoring a single-valued steady branch.

A dedicated follow-up study will connect this closure-level condition to radiation-MHD simulations by measuring the response of $W_{R\phi}/(P_{\rm gas}+P_{\rm rad})$ to $X$, calibrating the resulting $\alpha_{\rm eff}(X)$, and applying it to disk-variability predictions.

Moreover, the $\alpha_{\rm x}$ closure smooths the scale-height response of the inner disk while raising $\Sigma$ and $\tau$, suggesting that moderate-to-high $\dot{m}$ AGN disks can remain geometrically thin and more optically thick without immediately becoming slim. By lengthening thermal and inflow timescales, it provides a natural route to accretion-state dependent variability without invoking large-amplitude RPD limit cycles, and may affect irradiation geometry, BLR illumination, wind launching, and continuum-size diagnostics in luminous AGNs.

\begin{acknowledgements}
This work was supported by the F.R.S. FNRS under the research grant IISN 4.4503.19. DH is F.R.S.-FNRS Research Director.
\end{acknowledgements}

\bibliographystyle{aa}
\bibliography{naddaf}

\end{document}